\documentclass[12pt]{article}
\usepackage{amssymb}
\usepackage{amsmath}
\def\dd{{\rm d}}\def\ee{{\rm e}}\def\ii{{\rm i}}

\def\cut{\hfil\break}
\def\beq{\begin{equation}}\def\eeq{\end{equation}}
\def\bea{\begin{eqnarray}}\def\eea{\end{eqnarray}}

\begin{document}
 
\title{Incorporating particle creation and annihilation into Bohm's Pilot Wave model}
 
\author{Roman Sverdlov
\\Raman Research Institute,
\\C.V. Raman Avenue, Sadashivanagar, Bangalore -- 560080, India}
\date{May 12, 2010}
\maketitle
 
\begin{abstract}

\noindent The purpose of this paper is to come up with a Pilot Wave model of quantum field theory that incorporates particle creation and annihilation without sacrificing causality. In some sense, this work echoes the work of Nikoli\'c in \cite{nikolicb} and \cite{nikolicf} (I call "visibility" what he calls "effectiveness"), but in my work I choose position and visibility as beables, as opposed to field beables that are used in his. 

\end{abstract}
 
\subsection*{1. Introduction}

The problem of the interpretation of quantum mechanics is well known. While its predictions are formulated in terms of the probabilities of the collapse of wave function after its interaction with a so-called ``measuring device", the definition of ``measurement" as well as what happens while it is carried out is not understood. Furthermore, in order for the classical system called ``measuring device" to exist in the first place, that multi-particle system has to somehow ``collapse" without any a priori measuring device to cause that collapse, which leaves us with a circular argument. Finally, the very concept that the result of a measurement can not be predicted with absolute certainty violates our deep held beliefs in determinism.

These question were addressed in the context of non-relativistic quantum mechanics by Bohm, and subsequently by the follow-up work of others. According to these models, particles and waves exist as separate substances.  Waves evolve according to Schr\"odinger's equation, while particles are moving according to well defined trajectories based on the interaction with that wave, which is called \emph{guidance equation}. Both of these evolution equations are completely deterministic. Bohm had shown that if we take the point of view of configuration space rather than our usual one, then the model predicts the probability distribution of $\vert \psi \vert^2$ as well as the appearance of a collapse of the wave function. That collapse is not real, since the same dynamics governs the system at all times. Thus, Bohm called it ``effective collapse". Furthermore, $\vert \psi \vert^2$ is a probability in a classical sense rather than quantum mechanical: the system itself is deterministic, and the use of probabilities is only necessary due to our own lack of detailed knowledge of the initial conditions of the system. 

 However, this theory is not compatible with creation and annihilation of particles. After all, these processes are not continuous and, therefore, can not be construed as part of the continuous equation of the Pilot Wave model. In \cite{jumps} a theory was proposed according to which a continuous evolution is being interrupted by these jumps. But the timing of these jumps was chosen at random, which violates determinism. 

The goal of this paper is to make these jumps ``continuous" which will allow them to be parts of a deterministic model. I do that by saying that point particles don't truly disappear, but rather they become ``invisible". This allows me to postulate ``visibility degrees" that vary \emph{continuously} between $0$ and $1$.  While the particles are distinguishable and none of them are created or destroyed, the particles with visibility degree $0$ are not subject to interaction, which leads to the \emph{appearance} of the particles' annihilation. Then, when its visibility degree stops being $0$, it leads to the \emph{appearance} of particle creation. 

It is important to point out that "less visible" particles are not "less real". On the contrary, \emph{all} particles have well defined positions in space \emph{at all times}. Visibility is simply the degree of interaction of a particle, \emph{not} a degree of its reality. For example, if the only interaction that exists was electromagnetic one, the neutrons would have been "invisible"; but they would still be real. In our case, due to the fact that Pilot Wave model is Aristotelian, an interaction of a particle is reflected in its velocity rather than acceleration. Therefore, the "invisible" particles are almost stationary in space, as they await something to make them "visible" at which point they will be allowed to move. 

In general, the greater their visibility degree is, the stronger particles interact. Thus, a particle with visibility degree 1/2 appears as half a particle. However, the particle spends very little time in such state (although still a finite amount of time), thus \emph{most of the time} the number of particles is integer.  I accomplish this by introducing compactified extra dimension, $x_4$, and claiming that the visibility degree of a particle is given by $f(x_4)$ for some specified differentiable function $f$. If $f$ happens to be a differentiable approximation to a step function, this would give me the desired result.

The Hamiltonian is the same as in ordinary quantum field theory, which means that creation and annihilation operators represent \emph{instantaneous} creation and annihilation, as usual. Accordingly, this Hamiltonian only determines the probability amplitudes of the states where all particles have visibility degrees $0$ or $1$. However, the probability amplitude is then extrapolated to states where point particles have other visibility degrees through weighted averages of probability amplitudes of the above \emph{extreme} states, where the products of visibility degrees serve as coefficients.  We then use the probability amplitude, as opposed to the Hamiltonian, to write down a Pilot Wave model of particles with continuously changing positions and visibility degrees. 

Of course, as with all Pilot Wave models, this proposal violates the principles of relativity at its core, by defining beable as a non-local \emph{spatial} configuration of particles that evolves in time, the latter being a \emph{preferred} time direction.  While we know that creation and annihilation of particles is linked to relativity, according to this model this is purely a coincidence. Likewise, the fact that the Hamiltonian is taken from \emph{relativistic} quantum field theory is viewed as coincidental, as well.  Nevertheless, the fact that it \emph{happened} to borrow both of these features allows us to hope that it predicts the \emph{appearance} of relativity in the lab. 

Finally, it is important to mention that similar idea has been proposed by Nikoli\'c back in 2004 (see \cite{nikolicb} and \cite{nikolicf}): what I call "visibility" of a particle, he calls "effectiveness". Unfortunately, I was unaware of it when I posted first two versions of this paper, and I am apologizing to Nikoli\'c for not having given him appropriate references until now. There are, however, some differences between my work and his. I will compare and contrast these concepts in Chapter 6 of the current version. 

\subsection*{2. Pilot Wave model for fixed number of particles}

In this section we will briefly outline Pilot Wave model originally proposed by Bohm in 1952. For more details, the reader is referred to his original work in \cite{Bohm1}, which is continued in \cite{Bohm2}. 

If the total number of particles in the universe is fixed (say, it is $N$), it is well known that a configuration of \emph{all} of them is represented by a single point in $3N$-dimensional \emph{configuration space}.  It is conjectured that the defining feature of so-called "classical" systems is that they are complex enough to "remember" their interaction with their environment. Their "memory", of course, corresponds to configuration of their particles and, therefore, can be encoded in a single point in a configuration space (the latter represents all particles in the universe, \emph{including} the "classical" system we are interested in). 

Therefore, during double slit experiment the screen "remembers" where it was hit by an electron and when. Thus, the points in a configuration space corresponding to the memory of an electron hitting a screen at a given location and at a given time are not accessible from the other points corresponding to the memories other outcomes. In other words, the wave function in configuration space, that represents all the particles in the universe, splits into several non-overlapping branches, each corresponding to one of the outcomes of double slit experiment. 

According to Everett, each of these branches represents one of the "parallel universes". Bohm, on the other hand, believes that there is one \emph{single} universe which is defined \emph{not} as one of the branches, but rather by a particle in configuration space (or, equivalently, $N$ particles in ordinary one) that is localized \emph{at all times}. Its behavior is described by \emph{guidance equation},
\beq \frac{d\vec{x} (t)}{dt} = \frac{1}{m} Im \; (\vec{\nabla} \psi )(\vec{x} (t)). \eeq
He have shown that, after sufficient time has passed,  the probability density of finding that particle can be, indeed, approximated by $\vert \psi \vert^2$,  provided that $\psi$ evolves according to Schr\"odinger's equation. At the same time, after the branches split, the particle will be forced to stay in only \emph{one} of these branches. Thus, there will be an appearance of collapse of wave function, despite the fact that no such collapse occurred. This is called "effective collapse".

\subsection*{3. Creation and annihilation of particles}

We are now ready to move on to introduce creation and annihilation of particles into the Pilot Wave model. Among other things, this means replacing the wave function described by Schr\"odinger's equation by probability amplitudes described by quantum field theory. These probability amplitudes describe states with varying numbers of particles. Our goal is to describe a way in which one multiparticle state evolves into the other deterministically, while the number of particles is not fixed. 

As was explained in the introduction, the key concept that allows us to introduce creation and annihilation is introducing the notion of \emph{visibility} of particles, which continuously varies between $0$ and $1$. Thus, the total number of degrees of freedom is fixed at all times, and while particles with visibility $0$ do not interact, they continue to exist.  Furthermore, visibility can only change in a continuous fashion. Thus, while it is near $0$ or $1$ \emph{most of the time}, it takes a small but finite amount of time to make a transition between these two states. This allows us to incorporate particle creation and annihilation into Pilot Wave model, since now these processes are continuous and thus can be described in terms of a differential equation.

In order to formally introduce the notion of visibility of the particle, we introduce an extra coordinate, $x_4$, that is compactified: $x_4 + 2 \pi = x_4$ (not to be confused with the time coordinate; we agree that the time coordinate is $x_0$, while $x_4$ is an extra spacelike one; in some literature it is denoted by $x_5$ to avoid confusion, but in this paper we stick with calling it $x_4$). We make sure that visibility is near $0$ or $1$ most of the time by setting it to be $f(x_4)$ rather than $x_4$ itself, and we choose a fixed function $f$ in such a way that $1- \epsilon < f(x_4) <1 $ whenever $x_4 \in (\delta; \pi - \delta)$, and $0< f(x_4) < \epsilon$ whenever $x_4 \in (\pi + \delta, 2 \pi - \delta)$, for sufficiently small $\epsilon$ and $\delta$. The specific way in which we choose that function is not important, as long as that function is continuous and differentiable. For definiteness, we can set
\beq
f(x_4) = \frac{1}{2} + \frac{1}{\pi} \tan^{-1} (n\, \sin\, x_4)\;,
\eeq
where $n$ is some fixed very large number. 

However, as mentioned in the introduction, the Hamiltonian continues to be the one of ordinary quantum field theory. In particular, the creation and annihilation operators of the Hamiltonian are \emph{instantaneous}, the particles living in the state space in which it acts only have visibilities \emph{exactly} zero or one, and finally, due to unitarity, the number of particles can be arbitrarily large, and thus exceed the number of degrees of freedom of the space we are interested in. The approach of this theory is to come up with ``projections" of probabilities from that space, onto the space we are interested in. The former is defined directly based on the amplitudes computed based on the rules of quantum field theory, the latter is not. 

Let's first consider a toy model, where we have only one spin-0 scalar field, and thus only one kind of particle. The space that is subject to standard quantum field theory is $Q = \mathbb{R}^0 \bigcup \mathbb{R}^3 \bigcup . . . \bigcup \mathbb{R}^{3n} \bigcup. . .$. We can denote a state in $Q$ where there are $n$ particles, at locations $\vec{x}_1^{(3)} \in \mathbb{R}^3$ through $\vec{x}_n^{(3)} \in \mathbb{R}^3$ by $\vert \vec{x}_1^{(3)}, . . . , \vec{x}_n^{(3)} \rangle$, where the (3) at the top serves to remind us that we are living in three-dimensional space, and do \emph{not} include the extra coordinate we were talking about earlier. Furthermore, in that three-dimensional space we have an ordinary quantum mechanical state, evolving by the usual rules of quantum field theory, 
\beq
\vert S(t) \rangle = \ee^{\ii H(t-t_0)} \vert S(t_0) \rangle\;,
\eeq
where $H$ is a Hamiltonian defined in the usual way: with creation and annihilation operators satisfying prescribed commutation (or anti-commutation if we replace bosons with fermions) relations. The Hamiltonian is three-dimensional. In other words, it takes the form
\beq
H(t) = \int \dd^3 x^{(3)} H(\phi(\vec{x}^{(3)}))\;.
\eeq
where $\phi (\vec{x}^{(3)})$ is a field operator at $x^{(3)}$. We can now view the quantum field theory amplitude as a function $\psi^{(3)} \colon Q \rightarrow \mathbb{C}$ (there is an unfortunate lack of letters in the alphabet, so we should remember not to confuse $\psi$ with the fermionic field). That amplitude is defined as
\beq
\psi^{(3)}(\vec{x}_1^{(3)}, . . . , \vec{x}_n^{(3)})
= \langle\vec{x}_1^{(3)}, . . . , \vec{x}_n^{(3)} \vert \ee^{\ii H(t-t_0)} \vert s(t_0)\rangle
\eeq
where $\phi$ is field operator. However, our beables are supposed to live in $4+1$ dimensions rather than $3+1$. Finally, for convenience, we will define a so-called "probability density" on $Q$:
\beq \rho^{(3)} (\vec{x}_1^{(3)}, . . . , \vec{x}_n^{(3)}) = \vert \psi^{(3)}(\vec{x}_1^{(3)}, . . . , \vec{x}_n^{(3)}) \vert^2 \eeq
This, however, is not the actual probability density since the beables live in $B$, not $Q$. Therefore, we have to "convert" $\rho^{(3)} \colon Q \rightarrow \mathbb{R}$ to $\rho^{(4)} \colon B \rightarrow \mathbb{R}$.  We will do that by ``projecting" the states in $4+1$ dimensional space onto the states in $3+1$ dimensional ones, and taking weighted averages of the values of $\rho^{(3)}$, to come up with a function $\rho^{(4)} \colon B \rightarrow \mathbb{R}$.

In order not to have ambiguities regarding the order of the coordinates in our configuration space, we can formally let the particles be distinguishable. This can be done without violating the statistics of indistinguishable particles. Consider, for example, a toy model of two particles which are only allowed to occupy two possible states. According to standard theory the system has only three possible states: both particles in state $1$, both particles in state $2$, and one particle in each state. They can be assumed to have probability $1/3$ each. In our case, however, we can formally break the last state we mentioned into two sub-states. Thus, we have four states, with probabilities $1/3$, $1/3$, $1/6$ and $1/6$. This is still consistent with the statistics of quantum mechanics.

Now that we have done that, we can formally define an operation $P \colon Q \rightarrow Q$ that ``gets rid" of the particles with a specified numbers. More precisely, if $S \subset \mathbb{N}$, then the only particles that are being \emph{retained} are the ones whose numbers are elements of $S$. For example, if $S = \{2, 5, 7\}$, then $P_S (x_1, . . . , x_8) = (x_2, x_5, x_7)$. We are now ready to define a probability density, $\rho^{(4)}$, in the space in which our beables live. If the number of beables is $N$, then their configuration space is $(\mathbb{R}^3 \times \Gamma)^N$, where $\Gamma$ is an extra dimension that we need to use in order to define visibility. The probability density $\rho^{(4)}$ in that space is defined as
\beq \rho^{(4)} (x_{11}, . . ., x_{14}, . . ., x_{N1}, . . . , x_{N4}) = \eeq
\beq = \frac{1}{I} \sum_{S \subset \{1, . . . , N\}} \Big(\prod_{i \in S} f(x_{i4}) \Big)\, \Big(\prod_{j \not\in S} (1-f(x_{j4}))\, \rho^{(3)} (P_S (x_1^{(3)}, . . . , x_N^{(3)})\Big)  \nonumber \eeq
where
\beq I = \int d^4 x_1^{(4)} . . . d^4 x_N^{(4)} \vert \sum_{S \subset \{1, . . . , N \}} \Big( \prod_{i \in S} f(x_{i4}) \Big) \Big( \prod_{j \not\in S} (1 - f(x_{j4})) \rho^{(3)} (P_S (x_1^{(3)} , . . . , x_N^{(3)})), \eeq
\beq
x_k^{(3)} = (x_{k1},x_{k2},x_{k3})\;;\quad x_k^{(4)} = (x_{k1},x_{k2},x_{k3},x_{k4})
\eeq
and $x_{k4}$ is the extra coordinate that is being used for visibility. Here, it is understood that the total volume of our space is finite (for example, we might be living in a compact sphere and/or a space that has boundaries). Then, due to the finite number of particles, the volume of configuration space is finite as well, which is why the above expressions are well defined. Finally, in the above expression, the physics is contained in the numerator, while the denominator is just normalization.  

$\rho^{(3)} (P_S (x_1^{(3)}, . . . , x_N^{(3)}))$ represents the probability amplitude, predicted by quantum mechanics, with only the particles with indexes in $S$ being present. Thus, the sum over all possible $S \subset \{1, . . . , N \}$ of that represents the superposition of the probability densities associated with all states with number of particles less than or equal to $N$. However, different $S$ come up with different coefficients, which means that some contribute more than others. In particular, since most of the time $f(x_4)$ is either close to $0$ or $1$, the same applies to each of the above coefficients. Therefore, the situation can be broken into three cases: 

1) All particles whose numbers are elements of $S$ have visibility close to $1$. All other particles have visibility close to $0$

2) Some of the particles whose numbers are elements of $S$ have visibility close to $0$

3) Some of the particles whole numbers are \emph{not} elements of $S$ have visibility close to $1$. 

In case of the first option, the product of $f(x_k)$ over all $k \in S$ is approximately $1$. Likewise, the product of $1-f(x_k)$ over all $k$ \emph{not} in $S$ is also approximately equal to $1$. Thus, neither of the products have serious effect on the term. In case $2$, on the other hand, the product of $f(x_k)$ over $k \in S$ will be close to $0$, and thus the whole term will be close to $0$ as well. Finally in case $3$ the product of $1-f(x_k)$ over $k\not\in S$ will be close to $0$, which, again, would mean that a given term is close to $0$. Thus,  the sum can be approximated as 
\beq \sum_{S \subset \{1, . . . , N \}} \Big( \prod_{i \in S} f(x_{i4}) \Big) \Big( \prod_{j \not\in S} (1 - f(x_{j4})) \rho^{(3)} (P_S (x_1^{(3)} , . . . , x_N^{(3)}) \Big) \approx \rho^{(3)} (P_{T} (x_1^{(3)} , . . . , x_N^{(3)})) \eeq
where $T$ is a set of integers for which the visibilities of corresponding particles are close to $1$. The right hand side of the above equation, of course, is $\rho^{(3)}$ dictated from quantum field theory. At the same time, the above approximation is true only in \emph{most} cases but not all. After all, in light of continuity, sometimes visibilities are far away from both $0$ and $1$. The latter allows a continuous transition from one "typical" case to another. But, at the same time, it has no impact on our observations since, in light of the specifics of $f$, the transition region in $x_4$ is very small. 

By looking at the left-hand side, we note that we now have a wave function over $x_k^{(4)}$ rather than $x_k^{(3)}$, which means that we have taken the extra coordinate seriously. However, we have been using $\rho^{(3)}(x^{(3)})$,  in order to define $\rho^{(4)}(x^{(4)})$. Physically, we have used quantum field theory to compute the probability densities of \emph{extreme} states where particles have visibilities of either $0$ or $1$, and then we have used a weighted average to extrapolate it on other states. In the case of extreme states, we no longer need $x_4$ since its only purpose is a \emph{continuous} variation of visibility, which is why we have $x^{(3)}$ on our right-hand side. On the other hand, the coefficients of our weighted average are based on visibility, i.e., the fourth coordinate, which is why we have $x^{(4)}$ on our left-hand side, as well as the coefficients $f(x_{i4})$ and $1-f(x_{i4})$ at the right.   

In general, the probability density of a state where particles have visibilities between $0$ and $1$ is equal to that of a mixed state. However, this is still viewed as a \emph{pure} state.  The situation is somewhat similar to saying that the probability density of a \emph{pure} state consisting of half a particle is equal to the average of probability densities of vacuum and one-particle states. This, however, is not a perfect analogy either. In our case we \emph{do} have all the particles present, as a \emph{whole}. This is evidenced by the fact that the number of degrees of freedom of our configuration space is $4N$. However, the degree of interaction of some of the particles is weaker, which lead them to \emph{appear} as half a particle, or even no particle. This weaker interaction is expressed in a gradient of a wave function with respect to $x_k^{(3)}$ being smaller than the one for other particles. 

It is important to note that in the definition of $\psi^{(4)}$ we have only used the probability amplitudes of states where the number of particles is less than or equal to $N$. The validity of this depends on the conjecture that quantum field theory is formulated in such a way that the probability of generating more than $N$ particles is very small. If such is the case, we will not be able to disprove our theory in a lab. This would allow us to hypothesize that our theory is, in fact, exact rather than an approximation. What gives us hope is the fact that the existence of the upper bound on the number of particles is consistent with our everyday experience. However, explicit verification of this depends on appropriate renormalization procedure that assures us that probability amplitudes associated with multi-loop diagrams approach zero as the number of loops goes to infinity. The exploration of different kinds of renormalization techniques that can be used is beyond the scope of this paper.  

Finally, we are ready to introduce more than one field, and, accordingly, more than one type of particle. Up until now we have established that the total number of particles is fixed. Now that we will have more than one kind of field, $\phi_1, . . . , \phi_A$, we need a more detailed description of particle numbers: the number of particles of field $k$ is set to be $N_k$. The space in which beables live is given by 
\beq
B = (\mathbb{R}^3\times\Gamma)^{N_1} \times ...
\times (\mathbb{R}^3\times\Gamma)^{N_A}\;,
\eeq
while the quantum mechanical probabilities are defined on the space 
\beq
Q = \bigcup_{n_1, . . . , n_A} \mathbb{R}^{3n_1} \times . . . \times \mathbb{R}^{3n_A}\;.
\eeq
The probability amplitude $\psi^{(3)}$, as well as probability density $\rho^{(3)}$ continue to be defined in the usual way, except that now the state that is used to define it consist of more than one kind of particles, and finally the procedure for obtaining $\psi^{(4)} \colon B \rightarrow \mathbb{C}$ out of $\psi^{(3)} \colon Q \rightarrow \mathbb{C}$ is identical, up to the above replacements. 

It is important to note that, while we have done it for spin-0 bosons, the prescription is identical for bosons and fermions alike, and can be used for all spins. The only difference is in the Hamiltonian, where anticommuting operators will be used instead of commuting ones. Nevertheless, the usual methods of quantum field theory will still give us the probability amplitudes on Fock space, which is all we need. 

Since we do not regard spins as parameters of our configuration space, we get rid of them while taking projection $Q \rightarrow B$. In particular, while $\rho^{(3)}$ will be a function of spins, $\rho^{(4)}$ will not be, since the latter will involve the sum over all possible spin assignments. Nevertheless, since spins are still present in $Q$, in practice some of the spin assignments will have significantly greater contribution than others to the dynamics of $B$. Furthermore, the information about spins will be depicted in $B$ through the configuration of photons that were emitted by the particles that supposedly have these spins. Likewise, the information about the polarization of photons can be guessed from their configuration, as well as the configuration of particle-antiparticle pairs that they have produced.

Finally, it is important to say a few words about ``rotation" of one particle into the other. In light of the fact that the number of particles of each kind is specified, according to the framework of the theory, \emph{all} available particles exist at all times, and they never change from one kind to the other. What \emph{does} happen is that their visibility changes. Thus, if an electron's visibility changes from $1$ to $0$, while a neutrino's visibility changes from $0$ to $1$, and if these processes occur close enough in time, this would lead to the \emph{appearance} that the electron has ``rotated" into a neutrino. This, however, is not the case, since the given electron never stopped being an electron, and the given neutrino has never been anything else. The ``electron" that the neutrino has turned into is a completely separate particle. 

Now, the key to the appearance of ``rotation" of the electron into a neutrino is a ``coincidence in time": the visibility of the electron had decreased approximately when the visibility of the neutrino had increased.  This, of course, is not really a coincidence, but instead a consequence of the Hamiltonian. In particular, if the visibility of the neutrino increased, but the visibility of the electron did \emph{not} decrease, then the main contributor to our weighted average would come from a non-existing diagram of neutrino being created out of nothing. This means that the probability density of this would be nearly zero. As far as the Hamiltonian (or Feynman diagrams) are concerned, we \emph{do} have the rotational symmetries, which explains the ``coincidence". But since the space in which beables live in ($B$) is separate from  the one that is subject to quantum field theory ($Q$), there is no ``rotation" as far as $B$ is concerned. 

In principle, we could have used rotation as a replacement for visibility, since this, too, is a continuous operation, and in fact is more natural. We chose not to do that because we would like to develop a general framework for all quantum mechanical processes, and the decay of a photon into an electron and a positron can't be viewed as such. On the other hand, we \emph{could} have avoided the problem in case of photon emission, if we were to make the choice to only use position beables for fermions and field beables (see, for example, \cite{minimalist}) for bosons (which would basically be a different paper, since our current choice is to use position beables for both). Thus, the ultimate reason we need visibility is that pair creation is ``worse" than emission. This is due to a violation of crossing symmetry, which is linked to a violation of relativity in the definition of configuration space. 

\subsection*{4. Pilot Wave model}

\subsection*{4.1 The model}

We are now ready to introduce the Pilot Wave model. We would like to define a smooth vector field $\vec{v} \colon \mathbb{R} \times (\mathbb{R}^3 \times \Gamma)^N \rightarrow (\mathbb{R}^3 \times \Gamma)^N$, and then postulate an equation of motion for $\vec{x} \in (\mathbb{R}^3)^N$ to be
\beq
\frac{\dd\vec{x}}{\dd t} = \vec{v} (t, \vec{x}(t))\;.
\eeq
By the same argument as before, $\vec{v}$ should be a solution of the differential equation
\beq
\frac{\partial \rho^{(4)}}{\partial t} = - \vec{\nabla} \cdot (\rho^{(4)} \vec{v})\;,
\eeq
and this time $\vec{\nabla}$ is taken with respect to $(\mathbb{R}^3 \times \Gamma)^N$ coordinates. 

However, unlike the case discussed in the previous section, we no longer have the Schr\"odinger equation at our disposal, so we can not re-express it in a form similar to the one we obtained in the previous chapter. The only price to pay for this is that the solution to this equation is not unique. Therefore, we "fix a gauge" by postulating an additional equation that assures the uniqueness of the solution, namely 
\beq
\rho^{(4)} \vec{v} = \vec{\nabla} \lambda\;,
\eeq
for some scalar function $\lambda \colon (\mathbb{R}^3 \times \Gamma)^N \rightarrow \mathbb{R}$. This completely specifies $\vec{v}$:
\beq
\vec{v} = \frac{1}{\rho^{(4)}} \vec{\nabla} \lambda\;,
\eeq
where $\lambda$ is a solution to the $4N$-dimensional Laplace equation on a cylinder,
\beq
\nabla^2 \lambda = - \frac{\partial \rho^{(4)}}{\partial t}\;,
\eeq
with the boundary conditions that $\lambda$ is $0$ at infinity.  

If we want, we can write the above in the integral form. In order for this to be more transparent, we notice that the problem is identical to electrostatics if we replace $\partial \vert \psi \vert^2$ with $\sigma$, and interpret it as a charge density. Then, $\lambda$ can be interpreted as electric potential, and $\vec{\nabla} \lambda$ as an electric field. Thus, our guidance equation becomes
\beq \vec{v} = \frac{\vec{E}}{\rho^{(4)}}  \eeq
We now use the method of images in order to compute $\vec{E}$ in compactified geometry. We consider an imaginary situation where coordinates $x_{4k}$ extend to infinity, and charge density is periodic in these coordinates, 
\beq \sigma (\vec{x}) = \sigma (\vec{x} + 2 \pi \hat{x}_{4k}) \eeq
By translational symmetry, we notice that  
\beq \vec{E} (\vec{x}) = \vec{E} (\vec{x} + 2 \pi \hat{x}_{4k}) \eeq 
from which it is easy to see that the identical copy of $\vec{E}$ satisfies the same differential equation in the compactified geometry, without having any discontinuities at $x_{4k} = 2 \pi n$. Thus, this is a solution we are seeking. Now, from the non-compact case, we can show that the area of the sphere is given by  
\beq A = \frac{2 \pi^{2N}}{(2N-1)!} r^{4N-1}\eeq
This tells us that the electric field is 
\beq \vec{E} (\vec{x})= \frac{(2N-1)!}{2 \pi^{2N}} \sum_{a_1, . . . , a_N} \int d^{4N} x' \frac{ \sigma (\vec{x}' )(\vec{x}' - \vec{x} + \vec{R}_{a_1, . . . , a_N})}{ \vert \vec{x}' - \vec{x} + \vec{R}_{a_1, . . . , a_N} \vert^{4N}} \eeq
where
\beq \vec{R}_{a_1, . . . , a_N} = \sum 2 \pi a_i \hat{x}_{4i} \eeq
is a displacement of an "image charge". Now by substituting back 
\beq \vec{E} \rightarrow \rho^{(4)} \vec{v} \; ; \; \sigma \rightarrow \frac{\partial \vert \psi \vert^2}{\partial t} \eeq
we obtain a guidance equation
\beq \vec{v}(\vec{x}) = \frac{(2N-1)!}{2 \pi^{2N} \rho^{(4)}(\vec{x})} \frac{\partial}{\partial t} \sum_{a_1, . . . , a_N} \int d^{4N} x' \frac{ \rho^{(4)} (\vec{x}') (\vec{x}' - \vec{x} + \vec{R}_{a_1, . . . , a_N})}{ \vert \vec{x}' - \vec{x} + \vec{R}_{a_1, . . . , a_N} \vert^{4N}} \eeq

\subsection*{4.2 Concrete example}

Let us illustrate the way the model works by working out a concrete example. Suppose we have two particles, living in one dimensional space. For simplicity, we will assume there is no visibility coordinate. So both particles have visibility $1$ at all times. Now, suppose each particle is subject to its own potential, that does not interact with the other particle. As a result of these potentials, the probability densities of these two particles, $\rho_1$ and $\rho_2$, look like $\delta$-functions moving with constant velocities, which are defined as follows: 
\beq \rho_1 (x_1, t) = \delta (x_1 - a_1 - v_{10} t) \; ; \; \rho_2 (x_2, t) = \delta (x_2 - a_2 - v_{20} t) \eeq
where $x_1$ and $x_2$ are respective coordinates of these particles. The constants $v_{10}$ and $v_{20}$ represent the constant velocity of the guidence wave, and are not to be confused with $v_1$ and $v_2$, which are oscillating velocities of the point particles. 

Now, from the point of view of configuration space, these two particles can be described as one \emph{single} particle moving in two dimensions. Its coordinates are $\vec{r} = (x, y)$, where $x=x_1$ and $y=x_2$. The density distribution in this space is given by 
\beq \rho (\vec{r}', t) = \delta^2 (\vec{r}' - \vec{a} - \vec{v}_0 t) \eeq
where $\vec{v}_0 = (v_{10}, v_{20})$. Thus, the "charge density" is given by 
\beq \sigma (\vec{r}', t) = \frac{\partial}{\partial t} \delta^2 (\vec{r}' - \vec{a} - \vec{v}_0 t) \eeq
Now, according to our model, the velocity of the particle is irrelevant to the electric field it produces. The latter is a function \emph{only} of a position of particle \emph{at that time}, and is independent of its velocity. Thus, in our case the electric field is given by 
\beq \vec{E} = \frac{1}{2 \pi} \int d^2 r' \frac{\vec{r'}-\vec{r}}{\vert \vec{r}' - \vec{r} \vert^2}  \frac{\partial}{\partial t} \delta^2 (\vec{r}' - \vec{a} - \vec{v}_0 t ) \eeq
where the power of $3$ in the denominator was replaced by power of $2$ in light of the fact that we are in two dimensions. By pulling the time derivative out of the integral, we obtain
\beq \vec{E} = \frac{1}{2 \pi} \frac{\partial}{\partial t} \frac{\vec{a}+\vec{v}_0 t-\vec{r}}{\vert \vec{a} + \vec{v}_0 t - \vec{r} \vert^2}  \eeq
Physically, this represents an electric field of a dipole. Since a charge is a time derivative of a probability density, the charge at the "front" part of the delta function is positive, and at the "back" part is negative. After all, at the "front" part the probability changes from $0$ to $\infty$, while at the"back" part it changes back from the $\infty$ to $0$. More concretely, the charge distribution is 
\beq \sigma (\vec{r}', t) = - \vec{v}_0 \cdot \vec{\nabla}' \delta^2 (\vec{r}' - \vec{a} - \vec{v}_0 t), \eeq
where $\vec{\nabla}'$ is a gradient with respect to $\vec{r}'$-coordinates. The electric field of the dipole we just described is given by 
\beq \vec{E} = - \frac{1}{2 \pi} \int d^2 r' \frac{\vec{r'}-\vec{r}}{\vert \vec{r}' - \vec{r} \vert^2}  \vec{v}_0 \cdot \vec{\nabla}' \delta^2 (\vec{r}' - \vec{a} - \vec{v}_0 t )\eeq
By pulling $\vec{v} \cdot \vec{\nabla}'$ out of integral,  this becomes 
\beq \vec{E} = - \frac{1}{2 \pi} \vec{v}_0 \cdot \vec{\nabla}' \frac{\vec{r}' - \vec{r}}{\vert \vec{r}' - \vec{r} \vert^2} \Big\vert_{\vec{r}' = \vec{a} + \vec{v}_0 t} \eeq
It is easy to see that $\vec{\nabla}'$ can be replaced with $\vec{\nabla}$, if we change overall sign. Thus, 
\beq \vec{E} =  \frac{1}{2 \pi} \vec{v}_0 \cdot \vec{\nabla} \frac{\vec{r}' - \vec{r}}{\vert \vec{r}' - \vec{r} \vert^2} \Big\vert_{\vec{r}' = \vec{a} + \vec{v}_0 t} \eeq
Finally, substituting $\vec{a} + \vec{v}_0 t$ for $\vec{r}'$, we get
\beq \vec{E} =  \frac{1}{2 \pi} \vec{v} \cdot \vec{\nabla} \frac{\vec{a} + \vec{v}_0 t - \vec{r}}{\vert \vec{a} + \vec{v}_0 t - \vec{r} \vert^2}  \eeq
It is easy to see that this expression matches the one with time derivative. If we evaluate either of the two, we get the final expression for $\vec{E}$: 
\beq \vec{E} = \frac{1}{2 \pi \vert \vec{a} + \vec{v}_0 t - \vec{r} \vert^2} \Big( \vec{v}_0 - \frac{2 (\vec{a} + \vec{v}_0 t- \vec{r})((\vec{a}+ \vec{v}_0 t- \vec{r}) \cdot \vec{v}_0 )}{\vert \vec{a} + \vec{v}_0 t - \vec{r} \vert^2} \Big) \eeq
Now, the velocity field is $\vec{v} = \vec{E}/\rho$, where $\rho$ is moving $\delta$-function. Thus, we get 
\beq \vec{v}= \frac{1}{2 \pi \vert \vec{a} + \vec{v}_0 t - \vec{r} \vert^2 \delta^2 (\vec{a} + \vec{v}_0 t - \vec{r}) }\Big( \vec{v}_0 - \frac{2 (\vec{a} + \vec{v}_0 t- \vec{r})((\vec{a}+ \vec{v}_0 t- \vec{r}) \cdot \vec{v}_0 )}{\vert \vec{a} + \vec{v}_0 t - \vec{r} \vert^2} \Big)  \eeq
Now, if we transfer it to a moving frame, 
\beq \vec{s} = \vec{r} - \vec{a} - \vec{v}_0 t  \; ; \; \vec{u} = \vec{v} - \vec{v}_0 \eeq
then the equation becomes 
\beq \vec{u} = \frac{\vec{v}_0}{2 \pi s^2 \delta^2 (\vec{s})} - \frac{\vec{s} (\vec{s} \cdot \vec{v}_0)}{\pi s^4 \delta^2 (\vec{s})} - \vec{v}_0 \eeq
We would now like to find a shape of the curve traced by these field lines. Since most of that curve is taken up by the region \emph{away} from the origin of the $\delta$ function, the first two terms on right hand side are infinitely larger than any other terms. Thus, the curve is defined by equating the sum of tehse two terms to $0$.  By substituting 
\beq \vec{s} \cdot \vec{v}_0 = \vert \vec{s} \vert \vert \vec{v}_0 \vert cos \; \theta \eeq
the equation inside the brackets tells us 
\beq \frac{\partial \vec{s}}{\partial \tau} = \vec{v_0} - \frac{2 \vec{s}  \vert \vec{v}_0 \vert cos \; \theta}{\vert \vec{s} \vert} \eeq
For convenience, we will select coordinate system for which the direction of $\vec{v}_0$ is one of the axes. Since $x$ and $y$ are already defined (namely, they are coordinates of each particle), we have to use $x'$ and $y'$ instead. The latter are defined as a rotation of $x$ and $y$ through
\beq \begin{pmatrix} x' \\ y' \end{pmatrix} = \begin{pmatrix} cos \; \alpha & sin \; \alpha \\ - sin \; \alpha & cos \; \alpha \end{pmatrix} \begin{pmatrix} x \\ y \end{pmatrix},\eeq
where
\beq \alpha = tan^{-1} \frac{v_{20}}{v_{10}} \eeq
Furthermore, we will define $(\xi, \theta)$ polar coordinates via
\beq x' = \xi \; cos \; \theta \; ; \; y' = \xi \; sin \; \theta \eeq
In these coordinates, 
\beq v_{0 \xi} = \vert \vec{v}_0 \vert \; cos \; \theta \; ; \; v_{0 \; \theta} = - \vert \vec{v}_0  \vert  \; sin \; \theta \; ; \;  s_{\xi} = \vert \vec{s} \vert \; ; \; s_{\theta} = 0 \eeq
Substituting these into $d \xi / d \tau$ gives us
\beq \frac{d \xi}{d \tau} = \vert \vec{v}_0 \vert \; cos \; \theta - \frac{2 \xi \vert \vec{v}_0 \vert \; cos \; \theta}{\xi} = - \vert \vec{v}_0 \vert \; cos \; \theta \eeq
and the equation for $d \theta /d \tau$ is
\beq \frac{d \theta}{d \tau} = - \frac{\vert \vec{v}_0 \vert \; sin \; \theta}{\xi} \eeq
The latter equation allows us to express $\xi$ in terms of $\theta$:
\beq \xi = - \frac{\vert \vec{v}_0 \vert \; sin \; \theta}{d \theta / d \tau} \eeq
Substituting this expression for $\xi$ into the $d \xi / d \tau$ equation gives us 
\beq \frac{d}{d \tau} \frac{sin \; \theta}{d \theta / d \tau} = cos \; \theta \eeq
After evaluating right hand side, this becomes
\beq cos \; \theta = cos \; \theta - \frac{d^2 \theta / d \tau^2}{(d \theta / d \tau)^2} \; sin \; \theta. \eeq
This implies that
\beq \frac{d^2 \theta}{d \tau^2} = 0, \eeq
or, equivalently,
\beq \theta = \omega \tau \eeq 
for some $\omega$. Substituting this into the equation for $d \theta /d \tau$, this becomes
\beq \omega = - \frac{\vert \vec{v}_0 \vert \; sin \; \theta}{\xi} \eeq
which gives us 
\beq \xi = - \frac{\vert \vec{v}_0 \vert \; sin \; (\omega \tau)}{\omega} \eeq
Converting it back into $(x', y')$ coordinates, we get
\beq x' = \xi \; cos \; \theta = - \frac{ \vert \vec{v}_0 \vert }{\omega} \; sin  \; (\omega \tau) \; cos \; (\omega \tau) = - \frac{\vec{v}_0}{2 \omega} \; sin \; (2 \omega \tau) \eeq
\beq y' = \xi \; sin \; \theta = - \frac{\vert \vec{v}_0 \vert}{\omega}  \; sin^2 \; (\omega \tau) = - \frac{\vert \vec{v}_0 \vert}{2 \omega} + \frac{\vert \vec{v}_0 \vert}{2 \omega} \;  cos (2 \omega \tau) \eeq
This describes the equation of a circle of radius $\vert \vec{v}_0 \vert / 2 \omega$, with center located at $x'=0$ and $y'= - \vert \vec{v}_0 \vert / 2 \omega$. In light of the fact that $\omega$ can be both positive and negative, these circles can be displaced in both directions. Furthermore, since the magnitude of $\omega$ is arbitrary, every single point in our configuration space is intersected by at least one of these circles. Since $\omega$ changes from circle to circle, its value becomes a \emph{function} of the coordinates of the point that it, supposedly, intersects. In particular, by using
\beq \omega \tau = tan^{-1} \frac{y'}{x'} \eeq
and
\beq \xi = - \frac{\vert \vec{v}_0 \vert}{\omega} \; sin \; (\omega \tau) \eeq
\beq \omega = - \frac{\vert \vec{v}_0 \vert \; sin \; tan^{-1} \frac{y'}{x'}}{\sqrt{x'^2 + y'^2}} = - \frac{\vert \vec{v}_0 \vert y'}{\sqrt{x'^2 + y'^2}} \eeq
we obtain
\beq \omega = - \frac{\vert \vec{v}_0 \vert (y \; cos \; \alpha - x \; sin \; \alpha)}{x^2 + y^2} \eeq
Now, these circles can give us a dependence of a velocity on position, which is what we mean by a "guidance equation" in Pilot Wave model. In particular, for every point in space we should find a circle that intersects it, and then "read off" the velocity from that circle. If we differentiate $x'$ and $y'$ along any given circle, we get
\beq \frac{dx'}{d \tau} = - \vert \vec{v}_0 \vert \; cos \; (2 \omega \tau) = 2 \omega y' - \vert \vec{v}_0 \vert \eeq
and
\beq \frac{dy'}{d \tau} = - \vert \vec{v}_0 \vert \; sin \; (2 \omega \tau) = 2 \omega x' \eeq
In tensor form, this becomes
\beq \frac{d}{d \tau}  \begin{pmatrix} x' \\ y' \end{pmatrix} = \begin{pmatrix} 0 & - 2 \omega \\ 2 \omega & 0  \end{pmatrix} \begin{pmatrix} x' \\ y'  \end{pmatrix} - \begin{pmatrix} \vert \vec{v}_0 \vert \\ 0 \end{pmatrix} \eeq
Now, since our aim is to define the dynamics of our two particles, and their coordinates are given by $x$ and $y$, we should convert $(x', y')$ coordinate system back to $(x, y)$. It is easy to check that the "rotated" form of the above equation is 
\beq \frac{d}{d \tau}  \begin{pmatrix} x \\ y \end{pmatrix} = \begin{pmatrix} cos \; \alpha & - sin \; \alpha \\ sin \; \alpha & cos \; \alpha \end{pmatrix} \begin{pmatrix} 0 & -2 \omega \\ 2 \omega & 0 \end{pmatrix} \begin{pmatrix} cos \; \alpha &  sin \; \alpha \\ - sin \; \alpha & cos \; \alpha \end{pmatrix} \begin{pmatrix} x \\ y \end{pmatrix} - \nonumber \eeq
\beq - \begin{pmatrix} cos \; \alpha & - sin \; \alpha \\ sin \; \alpha & cos \; \alpha \end{pmatrix}  \begin{pmatrix} \vert \vec{v}_0 \vert \\ 0 \end{pmatrix} = \begin{pmatrix} -2 \omega y \\ 2 \omega x \end{pmatrix} - \begin{pmatrix} \vert \vec{v}_0 \vert \; cos \; \alpha \\  \vert \vec{v}_0 \vert \; cos \; \alpha \end{pmatrix} \eeq
Finally, by substituting 
\beq \omega = \frac{ \vert \vec{v}_0 \vert \; (x \; sin \; \alpha - y \; cos \; \alpha)}{x^2 - y^2} \eeq
this becomes
\beq \frac{dx}{d \tau} = - \frac{2 \vert \vec{v}_0 \vert (xy \; sin \; \alpha \; - y^2 \; cos \; \alpha)}{x^2 + y^2} - \vert \vec{v}_0 \vert \; cos \; \alpha \eeq
\beq \frac{dy}{d \tau} = - \frac{2 \vert \vec{v}_0 \vert (x^2 \; sin \; \alpha \; - xy \; cos \; \alpha)}{x^2 + y^2} - \vert \vec{v}_0 \vert \; sin \; \alpha \eeq
Since everything we have done is in configuration space, $\vec{v}_0$ is not physical. So, in our final answer we have to re-express it in terms of $v_{10}$ and $v_{20}$. We should do the same with $\lambda$. It is easy to check that 
\beq \vert \vec{v}_0 \vert = \sqrt{v_{10}^2 + v_{20}^2} \; ; \; sin \; \alpha = \frac{v_{20}}{\sqrt{v_{10}^2 + v_{20}^2}} \; ; \; cos \; \alpha = \frac{v_{10}}{\sqrt{v_{10}^2 + v_{20}^2}} \eeq
Substituting these gives us 
\beq \frac{dx_1}{d \tau} = - \frac{2 (x_1x_2v_{20} - x_2^2 v_{10})}{x_1^2 + x_2^2} - v_{10} \eeq
\beq \frac{dx_2}{d \tau} = - \frac{2  (x_1^2 v_{20} - x_1 x_2 v_{10})}{x_1^2 + x_2^2} - v_{20} \eeq
Here, we have replaced $(x, y)$ with $(x_1, x_2)$ since we have converted a single particle in configuration space back into two particles in the ordinary one. In the above equation $x_1$ and $x_2$ describe the velocity of particle beables, while $v_{10}$ and $v_{20}$ describe the velocity of a wave that \emph{happened} to be localized at all times only in our example. The fact that $dx_1/d \tau$ depends on $x_2$ and $v_{20}$ clearly implies non-locality. We will discuss our interpretation of non-locality in chapter 5. 

Now, the above parameter $\tau$ is not to be confused with $t$. In light of the fact that the probability density away from the $\delta$-function source is infinitely small, we expect the velocity $\vec{v}= \vec{E}/ \rho^{(4)}$ to be infinitely large. Thus, the particle spends infnitely small time on that circle and then returns to the origin of the $\delta$-function. However, it spends infinitely small time at the origin of the $\delta$-function as well, due to its infinitesimal size. Therefore, we have to compare one infinitesimal time to another, to find which is larger; in other words, at any given moment in time, where is the particle most likely be found? 

In order to answer the above question, we replace $\delta$-function with a finite probability distribution, 
\beq \delta^2 (\vec{x}) \rightarrow \frac{\beta}{\pi} \; e^{- \beta \vert \vec{x} - \vec{v}_0 t - \vec{a} \vert^2} \eeq
where $\beta$ is a very large number. The electric field produced by the above distribution is given by 
\beq \vec{E}= \frac{\partial \vec{D}}{\partial t},  \eeq
where $\vec{D}$ is some other field, which satisfies
\beq \vec{\nabla} \cdot \vec{D} = - \frac{\beta}{\pi} \; e^{- \beta \vert \vec{x} - \vec{v}_0 t - \vec{a} \vert^2} \eeq
To simplify the calculation, we define $r$ to be 
\beq \vec{r} = \vec{x} - \vec{v}_0 t - \vec{a}  \eeq
and $r = \vert \vec{r} \vert$. The $\vec{D}$-producting charge enclosed within the sphere of radius $r$ is given by 
\beq Q_D (r) = - \int_0^r \frac{\beta}{\pi} \; e^{- \beta r^2} 2 \pi r' dr' = - (1 - e^{- \beta r^2}) \eeq
Thus, the $\vec{D}$-field is 
\beq \vec{D} (r) = - \frac{1-e^{-\beta r^2}}{2 \pi r} \; \hat{r} \eeq
Substitutting for $\vec{r}$ we get
\beq \vec{D}  = - \frac{1}{2 \pi} \frac{(1- e^{- \beta \vert \vec{x} - \vec{v}_0 t - \vec{a} \vert^2})(\vec{x} - \vec{v}_0 t - \vec{a})}{\vert \vec{x} - \vec{v}_0 t - \vec{a} \vert^2} \eeq
which gives us
\beq \vec{E} = - \frac{1}{2 \pi} \frac{\partial}{\partial t} \frac{(1- e^{- \beta \vert \vec{x} - \vec{v}_0 t - \vec{a} \vert^2})(\vec{x} - \vec{v}_0 t - \vec{a})}{\vert \vec{x} - \vec{v}_0 t - \vec{a} \vert^2} \eeq
It is easy to see that $\partial/ \partial t$ can be replaced with $- \vec{v}_0 \cdot \vec{\nabla}$. Thus, the expression for $\vec{E}$ becomes
\beq \vec{E} = \frac{1}{2 \pi} (\vec{v}_0 \cdot \vec{\nabla}) \Big( \frac{1-e^{-\beta \vert \vec{r} \vert ^2}}{\vert \vec{r} \vert} \; \hat{r} \Big) \eeq
where we have again expressed it in a condensed form, using $\vec{r}$. After evaluating the derivative, this becomes
\beq \vec{E} = e^{- \beta \vert \vec{r} \vert^2} \Big(\frac{\beta}{\pi} \hat{r} (\vec{v}_0 \cdot \hat{r}) - \frac{1}{2 \pi} \frac{\vec{v}_0}{\vert \vec{r} \vert^2} + \frac{1}{\pi} \frac{\hat{r} (\hat{r} \cdot \vec{v}_0)}{\vert \vec{r} \vert^2} \Big) + \frac{1}{2 \pi} \frac{\vec{v}_0 - 2 \hat{r} (\hat{r} \cdot \vec{v}_0)}{\vert \vec{r} \vert^2} \eeq
In order to obtain $\vec{v}$, we have to divide the above expression by $\rho = (\beta/ \pi) e^{- \beta \vert \vec{r} \vert^2}$. In light of the fact that $\beta$ is large, this would give us, in the vicinity of the origin,
\beq \vec{v} \vert_{\vec{r} \approx \vec{0}} = \hat{r} (\vec{v}_0 \cdot \hat{r}) + 0 (1/ \beta) \eeq
In light of the cosine of the angle that comes with inner product, $\vert \hat{r} (\vec{v}_0 \cdot \hat{r}) \vert < \vert \vec{v}_0 \vert$. Since the source of $\delta$-function moves right with a velocity $\vec{v}_0$, the particle "falls behind" within a time interval proportional to the width of the $\delta$-function. By dimensional analysis, the width of the $\delta$-function is proportional to $1/\sqrt{\beta}$; thus, the time it takes for the particle to "fall behind" is proportional to the same.Once the particle "fell behind" sufficienty far, its distance from the source of the $\delta$-function becomes sufficiently large for the earlier approximation to become valid. Thus, the particle starts on a large circle and then once it completes that circle it comes back to the $\delta$-function source \emph{from the front}.

If we consider \emph{only} the part of the circle where the distance to the source is greater than $\gamma$, then the probability density is smaller than $(\pi / \gamma) e^{-\beta \gamma^2}$. Thus, the velocity is bounded \emph{below} by something proportional to $\gamma e^{\beta \gamma^2}$, and the time spent in that region is bounded \emph{above} by something proportional to $(1/\gamma) e^{- \beta \gamma^2}$. Now, suppose $\gamma$ is a fixed small number, while $\beta$ goes to infinity. Then the above goes to $0$ \emph{faster} than $\sqrt{\beta}$. In other words, the time spent on circle minus $\gamma$-segment is \emph{much smaller} than the time spent in the origin of the $\delta$-function.

Now there is a mismatch: $\gamma$-segment is much larger than the origin of a $\delta$-function. We can adress this as follows. Since $\gamma$-segment is small enough, we can simply assume that whenever the particle is on a $\gamma$-segment, it is "very close to the origin", regardless of whether or not it is inside the $\delta$-function. Thus, we are only interested in comparing the time particle spends inside the $\gamma$-segment with the time it spends outside. 

Now, since the vicinity of $\delta$-function is part of $\gamma$-segment, the time the particle spends on $\gamma$-segment is \emph{greater} than the time it spends in the vicinity of $\delta$-function. Since the latter is of the order $1/\sqrt{\beta}$, the time the particle spends inside the $\gamma$ segment is either of the order $1/\sqrt{\beta}$ \emph{or greater}. On the other hand, as we said before, the time the particle spends \emph{outside} of that segment is proportional to $(1/\gamma) e^{- \beta \gamma^2}$, which is much smaller then this. Thus, it spends most of the time inside the $\gamma$-segment. Since $\gamma$ can be arbitrary small, we conclude that the particle spends most of the time arbitrary close to the origin of $\delta$-function. 

The other thing to take into account is that the radius of the circle is not fixed, which means that it can be arbitrary large, which would increase the time particle spends there. This point can be adressed as follows. The radius of a circle is determined by the angle by which it leaves the vecinity of $\delta$-function; in order for the former to be infinite, it has to leave the $\delta$ function in the direction \emph{exactly} opposite to the velocity. This, of course, has probability $0$. Thus, for any given $\epsilon$ we can say that the probability is greater than $1- \epsilon$ that the radius of the circle is bounded above by $f (\epsilon)$. If we view $\epsilon$ as finite, we can again use the finite radius of the circle, and repeat above argument. 

The fact that the particle spends most of the time close to origin of $\delta$-function is what we have expected. After all, we have \emph{postulated} that probability is "large" at that point and $0$ everywhere else. So if we got anything other than the above something would have been terribly wrong. The above example also shows a mechanism by which the desired probability was enforced: since velocity is inversely proportional to probability density, the particle "flies faster" past the region where probability density is small, thus it spends very little time there. In above example, the radius of a circle might be quite large. Thus, the particle does \emph{not} avoid the "forbidden" region. It simply moves very fast on that circle so that it is hard to "catch" it while its there. 

The fact that the particle is almost certain to be found in arbitrary small region does \emph{not} violate Heisenberg's uncertainty principle. After all, the fact that probability looks like $\delta$-function implies that it has Fourier components of very different frequencies; in other words, the uncertainty of momentum is very large. The velocity, of course, is fixed; but it is not to be confused with momentum. Rather, $m \vec{v}$ is merely some form of "average" of the momentum over a very wide bell curve. This interpretation is confirmed by a well known theory of wave packets: the momentum is defined in terms of a wave length; the velocity of a wave packet simply \emph{happens} to approximate the momentum but it can \emph{not} serve as a definition of one. 

\subsection*{4.3 Non-relativistic case}

It is important to mention that, while the model of section 4.1 is designed to reproduce relativistic quantum field theory, it also reproduces the results of non-relativistic quantum mechanics, by default. The reason it is not as explicit as in Bohm's case is that, from the point of view of quantum field theory, any particle emits and re-absorbs other particles, which makes the theory fundamentally different. The transition from quantum field theory to quantum mechanics occurs on larger time scales, since these are the scales in which speed of light "looks" infinite. Physically, the emitted particles "had time" to be reabsorbed on these larger scales. On the other hand, uncertainty principle demands that we have more and more emitted particles on the smaller ones.

Thus, it is not completely correct to view creation and annihilation of particles as "interrupting" otherwise-continuous process (although this research direction is still worth considering; and it was in fact developed by D\"urr at el in \cite{jumps} and other papers). On the contrary, since the smaller time scales are "building blocks" of the larger ones, fundamentally, as far as Hilbert space $Q$ is concerned, the process is not continuous at all, and the continuity is only an appearance on larger scales. Of course, in this paper we have restored continuity by introducing another space $B$. But in our case, the beables in $B$ "look" at the discontinuous processes in $Q$ and try to "mimic" them in "continuous" fashion, thanks to the continuous function $f$ we have introduced. This is very different from Bohm where continuity arises naturally.  That is why Bohm's argument for $\rho = \vert \psi \vert^2$ can not be explicitly reproduced. 

In order to explicitly reproduce non-relativistic quantum mechanics, we have to do very large number of Feynman diagrams (and show how each diagram is "reflected" in our beable space). The number of Feynman diagrams should be large enough to "transition" from perturbative case to the non-perturbative one. While it is conceivable to do that in principle, it is beyond our practical capacities. Problems like this have not been resolved yet; in fact this is a main reason why the exact solution for relativistic hydrogen atom has not yet been found.  On the other hand, general proofs that quantum mechanics arises as a low energy limit of quantum field theory do exist. The latter allows us to conclude that our model reproduced $\rho = \vert \psi \vert^2$ by default, without having to verify it explicitly.  

\subsection*{5. Non-locality of the theory}

\subsection*{5.1 Does non-locality allow us to reproduce QFT}

The Pilot Wave model we have presented has one unpleasant feature: the so-called "Coulumbs Law" we have been using is non-local.While all Pilot Wave models, by invoking the notion of configuration space, are non-local by default, the one at hand is a level more non-local. In particular, other Pilot Wave models are non-local from the perspective of ordinary space $\mathbb{R}^3$, while they are local from the perspective of $\mathbb{R}^{3N}$. On the other hand, this model is non-local from the points of view of both spaces. 

Let's, for example, take ordinary Bohm's Pilot Wave model. In that model, at any given point $\vec{x} \in \mathbb{R}^{3N}$ the velocity field $\vec{v} (\vec{x})$ is a function only of $\psi$ and $\partial \psi$ \emph{at} $\vec{x}$. The non-locality comes in the picture only if we "translate" $\vec{x}$ from $\mathbb{R}^{3N}$ to $\mathbb{R}^3$. In the latter case, different components of \emph{the same} point $\vec{x} \in \mathbb{R}^{3N}$ are interpreted as coordinates of \emph{different} points in $\mathbb{R}^3$, that exist "simultaneously", which leads to non-locality. This, however, does not change the fact that, as far as $\mathbb{R}^{3N}$ is concerned, everything is local.

On the other hand, in our case, things are non-local in $\mathbb{R}^{3N}$ itself, since that is the space in which we have used "Coulumb's Law", and, unlike electrodynamics, Coulumb's Law is non-local. The difference in locality in $\mathbb{R}^{3N}$ is reflected in $\mathbb{R}^3$ as well. In particular, if we have two non-interacting systems then, in Bohm's case, the behavior of the particles of one is completely independent of the other one. In our case this is no longer true.

First consider Bohm's case.  If we denote the components of $\vec{x} \in \mathbb{R}^{3N}$ corresponding to two non-entangled systems, then the wave function can be expressed as $\psi (\vec{x}) = \psi_1 (\vec{x}_1) \psi_2 (\vec{x}_2)$. Since logarithm is additive, this means that 
\beq ln \; \psi (\vec{x}) = ln \; \psi_1 (\vec{x}_1) + ln \; \psi_2 (\vec{x}_2) \eeq
This implies that 
\beq \vec{v}_1 = \frac{1}{m} \vec{\nabla}_1 \; Im \;  ln \; \psi (\vec{x}) = \frac{1}{m} \vec{\nabla}_1 \; Im \; ln \; \psi_1 (\vec{x}_1) \eeq
is completely independent of $\vec{x}_2$. Likewise, $\vec{v}_2$ is completely independent of $\vec{v}_1$. Thus, each non-interacting subsystem can be treated as if it is all there is in the universe, and we will still obtain the correct result.

Unfortunately, this is no longer true in our case. Consider, for example, one dimensional case in which there are only two particles. Furthermore, to make it even simpler, suppose there is no visibility coordinate. Thus, both particles have visibility $1$ at all times. Suppose the wave function corresponding to these particles are non-interacting wave packets, $\psi_1 (t, x- a - u_1 t)$ and $\psi_2 (t, x - b - u_2 t)$. In this case the configuration space looks like a plane and, as usual, there is only \emph{one} particle living in that space. The wave function that guides that particle is a wave packet 
\beq \psi (t, x, y) = \psi_1 (t, x- a - u_1 t) \psi_2 (t, y- b - u_2 t). \eeq
At any given $t$, this wave packet is centered around $(a + u_1 t, b + u_2 t)$. The "charge density" associated with this wave packet is 
\beq \rho = \frac{\partial \psi_1 (t, x -a - u_1 t)}{\partial t} \psi_2 (t, y- b - u_2 t)+ \frac{\partial \psi_2 (t, x -a - u_1 t)}{\partial t} \psi_1 (t, y- b - u_2 t) \eeq
Thus, the electric field associated with this is given by 

Thus, the "electric field" produced by that wave packet is proportional to the projections on the direction pointing at the nearby locations of this point. It is easy to see that a projection is a non-linear function of $x$ and $y$. Since velocity is proportional to the electric field, $v_x$ depends on $x$ \emph{and} $y$ in non-linear fashion. But from the point of view of ordinary space, $x$ signifies first particle and $y$ signifies the second one. Thus, the velocity of the first particle depends on the positions of \emph{both} of them, despite the fact that wave packets are not interacting with each other!

This, of course, raises a question. In the previous section we have shown that we have specifically designed our equation in such a way that the predictions of standard quantum field theory are reproduced. Yet, standard QFT does \emph{not} predict the above-described interaction of non-entangled systems. In order to answer this question, we have to consider a situation where a system has been measured \emph{twice}. First time we measured the location of the \emph{first} particle, and the second time we have measured the location of the \emph{second} one. We will observe non-local interaction between these particles if and only if there is a relation between outcomes of these two measurements.

As we have previously explained, there is no such thing as actual "measurement". Instead, the interaction with sufficiently complex system causes the wave function in configuration space to split into several branches. Thus, we need to introduce other particles, whose coordinates are $z_1, . . . , z_{N-2}$ which interact with our two particle system at two instances. After the first interaction, which occurs at $t=t_1$, the wave function splits into $M$ branches:
\beq \psi (t; x, y, z_1, . . . , z_{N-2}) \approx \sum_{k=1}^M \psi_k (t; x, y, z_1, . . . , z_{N-2}) \; , \; t>t_1 \eeq
and then after the second interaction, which occurs at $t=t_2$, the branch number $k$ breaks into $L_k$ sub-branches:
\beq \psi_k (x, y, z_1, . . . , z_{N-2}) \approx \sum_{l=1}^{L_k} \psi_{kl} (x, y, z_1, . . . , z_{N-2}). \eeq
Each of these times, our particle is found in only one of the sub-branches. Now, let $\delta x$, $\delta y$ and $\delta z_k$ be the displacements of the particle \emph{from the center of the branch that it occupies} when $t<t_1$. Let $(\delta x)'$, $(\delta y)'$ and $(\delta z_k)'$ be the displacements of the particle from the branch it occupies at $t_1 <t < t_2$. And, finally, let $(\delta x)''$, $(\delta y)''$ and $(\delta z_k)''$ be the displacement of a particle from its sub-branch at $t>t_2$.

 In other words, before the first split, the coordinates of the particle are $x_p$ and $y_p$, while the coordinates of the center of the branch that it occupies are $x_b$ and $y_b$. We let $\delta x = x_p - x_b$ and $\delta y = y_p - y_b$. After the first split, the coordinates of the particle are $x_p'$ and $y_p'$, while the coordinates of the center of a \emph{sub-branch} it occupies are $x_b'$ and $y_b'$. Then $(\delta x)' = x_p' - x_b'$ and $(\delta y)' = y_p'-y_b'$. Finally, after the second split the coordinates of the particle are $x_p''$ and $y_p''$ and the coordinate of a center of sub-sub-branch that it occupies are $x_b''$ and $y_b''$. And we define $(\delta x)'' = x_p'' - x_b''$ and $(\delta y)'' = y_p''-y_b''$.

 Now, suppose the times that these two splits occurred are $t_1$ and $t_2$. Since each split takes very little time, the position of a particle doesn't have time to change much during each of these splits. Thus, $x_p' (t_1 + \epsilon) \approx x_p (t_1 - \epsilon)$, and $x_p'' (t_2 + \epsilon) \approx x_p' (t_2 - \epsilon)$ (the same is true for $y$). However, the position of the center of wave packet does change a lot. After all, we are not even talking about the same wave packet: we are comparing the center of a wave packet before the split to the center of \emph{one of the branches} after the split. Thus, $\vert x_b' (t_1 + \epsilon) - x_b ( t_1 - \epsilon) \vert >>0$ and $\vert x_b'' (t_2 + \epsilon) - x_b' ( t_2 - \epsilon) \vert >>0$ (same is true for $y$). This means that $\vert (\delta x)' (t_1 + \epsilon) - \delta x ( t_1 - \epsilon) \vert >> 0$ and $\vert (\delta x)'' (t_2 + \epsilon) - (\delta x)' ( t_2 - \epsilon) \vert >> 0$

What we want to find out is the relationship between the locations of the particle in a configuration space after the first and second measurements. This depends \emph{only} on $(\delta x)'$, $(\delta y)'$ and $(\delta z_k)'$. From the point of view of determinism, we can say that we only need to know these three values at $t=t_1+ \epsilon$. Now, since the first measurement was performed over $x$ and \emph{not} over $y$, this means that the projection of the support of $\psi_k$ onto $y$ axis at $t = t_1+ \epsilon$ is approximately the same as the projection of the support of $\psi$ at $t = t_1 - \epsilon$. Therefore, $(\delta y)' \approx \delta y$. 

Now, since we don't see the actual wave (we only see a particle) we don't have means of knowing the exact values of $\delta x$ and $\delta y$. At the same time, however, we are able to make a good guess. After all, based on some set of measurements that were performed before the so-called "first measurement" (that is, the ones not discussed here) we had some idea of what we expected the outcome of the first measurement to be. Thus, the displacement of its result from the expected outcome would give us $\delta x$ and $\delta y$.

 In our particular case, since the first measurement was performed over $x$ and \emph{not} over $y$, we \emph{only} know $\delta x$. At the same time, the very fact that measurement was performed over $x$ means that the width of the support of $\psi_k$ in the projection to $x$ axis is much smaller than the width of original $\psi$. Thus, $(\delta x)'$ has nothing to do with $\delta x$. Since the outcome of the second measurement is based on $(\delta x)'$, our knowledge of $\delta x$ is not useful. On the other hand, since the measurement of $y$ was \emph{not} performed, as we said before, $(\delta y)' \approx \delta y$. Thus, the value of $\delta y$ \emph{would} be useful \emph{if only} we knew it. But, unfortunately, we don't know it, since we didn't perform measurement of $y$.

To sum it up, the non-locality that we are worried about comes in a form of non-linear dependence of the second experiment on $(\delta x)'$ and $(\delta y)'$. In order for that non-locality to be detected, we need to gather some information about either of these two quantities from the result of the first experiment. As we have explained, we failed to do that. As a result, the non-locality can not be observed. This, of course, is not surprising. After all, from the setup, we expect our theory to be consistent with the predictions of ordinary QFT; the latter does not predict non-locality.

\subsection*{5.2 Does non-locality allow us to reproduce effective collapse}

So far we have shown that, despite the fact that the theory is highly non-local, the "local" predictions of QFT are successfully reproduced. There is, however, a different concern. In light of non-local "Coulumb" interaction, a particle can "jump" from one branch to another. Since the probability density between the branches is nearly zero, the "charge density" (which is defined as time derivative of probability density) is nearly zero as well. Therefore, nothing keeps a particle away from that region! The only reason the probability of finding it there is nearly zero is that its velocity is inversely proportional to $\rho$. Thus, whenever the particle is in that region it moves very fast, and thus spends very little time there. But, during that little time that it does spend in that region, it might as well fly from one branch to another, if it does so quickly enough.

Again we need to ask ourselves the same question as we asked before: our theory is designed in such a way that the predictions of standard quantum field theory are reproduced. Is this still the case? Surprisingly, even if the particles did make the jumps from one branch to the other, the answer would have been yes! Strictly speaking, we can not use our experience to claim that the above scenario never happened. After all, our memory is encoded in a configuration of the particles of our brain. The latter, of course, is a part of configuration space. Therefore, if a particle made a jump from the universe $k$ to the universe $l$, our subsequent memory will be consistent with universe $l$. Thus, we will forget everything that happened in universe $k$ and, at the same time, falsely remember the things that happened at the universe $l$.

 However, I find the above to be very unsatisfactory.  After all, the main agenda of Pilot Wave model is to recover the notion of determinism. Now, the latter would be meaningless if the notion of time was just an illusion (i.e. our memory). True: the above picture is still deterministic, since it includes real time \emph{on top} of our memory. But, in some respect, this amounts to philosophical inconsistency. Besides, it leaves us without any reason behind our choice of Bohm over Everett (the latter being equally deterministic). After all, if our particle can appear in \emph{any} of the branches, it makes these branches equivalent to each other, which is the case in Everett's model.

 Thankfully, there is at least one way of arguing that the above jumps are very unlikely. Since the branches of the wave function do not interact, the integral of probability density over each branch is conserved, up to very good approximation. Since the charge density is \emph{time derivative} of the probability density, the \emph{total} charge of each branch is $0$. Thus, the only possible reason for a jump is \emph{fluctuation} of charge within each branch. Thus, we would like to argue that the fluctuation of charge is not detectable due to the fact that branches are "far away" from each other. 

As was explained in introduction, the decoherence is a consequence of microscopic changes of complex system. Thus, there are good news and bad news: the bad news is that, in terms of \emph{each} of the coordinates, the two branches might, in fact, be close to each other. The good news, on the other hand, is that  \emph{so many} coordinates are different between the branches, that their overall distance might be large.  Suppose, for example, there are $N$ coordinates, and two points are exactly distance $\epsilon$ apart in projection to each of them,  and each then their overall distance  is 
\beq d = \sqrt{\sum_{k=1}^N \epsilon^2} = \epsilon \sqrt{N}. \eeq 
In this case, if $N >> 1/\epsilon^2$, then the two points  will be "far away", as desired. 

There is one more issue to address. Even if branches are far away from each other, if the width of each branch is comparable with the distance between them, they will still not look like a point to the particle situated in the other branch.   Now, the width of each branch, in projection to any given coordinate, is the uncertainty of the position of a particle in $\mathbb{R}^3$ represented by that particular coordinate of $\mathbb{R}^{3N}$. From physical intuition we know that the particles that are entangled in some macroscopic system typically have very small uncertainty in their position (except, of course, for examples such as electrons in a conductor). On the other hand, free particles typically have very large position uncertainty. 

This tells us that each branch of a wave function is similar to a large one-dimensional object in three dimensional space: it is narrow in some dimensions and wide in others. If we have very long one dimensional charge distribution, its field is inversely proportional to the distance. This remains true even if its length is infinite! Thus, if we have two of them (each represents one of the branches), then a particle interacts far more strongly with its own branch than the other one. Thus, even if from time to time it will be pulled by the charge fluctuations of the other branch, the charge fluctuations in its own branch will quickly pull the particle back.

We have to now be a little bit careful. As we mentioned previously, the fact that the total charge of each branch is $0$ should be instrumental in our argument. \emph{If} such was not the case, then the fact that charge is a derivative of probability demands that particles \emph{should} be flowing into the "positively charged" branch. So if our argument does not imply that, there is something wrong with it. As expected, our argument \emph{does} in fact imply this. Since the sum of the integrals of probabilities over \emph{all} branches is $1$, the derivative of that sum is $0$. Thus, the total charge of \emph{all} branches is $0$. If one of the branches is positive, it forces at least one of the other ones to be negative. This tell us that some of the electric field likes will flow from one of the branches into the other one. A particle, following these lines, will make a transition.

Now lets go back to our case. As we have previously stated, due to decoherence the branches do not interact. This makes total probability within \emph{each} of the branches constant, and, therefore, implies that total charge of \emph{each} branch is $0$. This implies that the electric field lines that go from any given branch will return to that branch. Thus, a particle will stay in that branch \emph{unless} one of these field lines "run into" another branch on its way back to the original one.  In order for the latter \emph{not} to occur, the typical length of charge variations within each branch has to be a lot smaller than the distance between branches. 

Now, as we said earlier, each branch has vastly different size in projection to different coordinates. We only need to worry about the ones in projection to which the branch is "long". As previously said, the latter  represents the behavior of free particles. Thus, in order for electric field lines to be reasonably short, every free particle needs to be confined to some isolated system in $\mathbb{R}^3$. This is consistent with our everyday experience: a particle is either confined to our lab, or earth's atmosphere, or galaxy, etc.  Thus, even if the distance between branches is much smaller than the size of each branch, the electric lines will \emph{not} be running into neighboring branch as long as their distance (that is $\epsilon \sqrt{N}$) is much larger than the typical size of a lab. This, of course is easy to accomplish. 

Now, of course, a lot of what was said in this section is just words. These "words" only imply that the jump is highly unlikely; but its probability is still non-zero. So one might still argue that if we "wait long enough" may be one day the jump will occur. However, the same objection can be raised against the standard, local, Bohm's theory as well. Since the value of wave function between branches is small but non-zero, it is possible that, one day, by some miracle the "local" guidance equation will carry a particle from one branch to another. So, in the defense of my work, I can say that it is not "worse" than the standard Bohm's theory. But, of course, both of these issues need to be investigated in a more quantitative manner. Unfortunately, it seems that due to the complexities of systems involved, the only means that are open to us are numeric; and, of course, numeric programs might require some few-particle simplifications which might distort what happens in the real world. These questions are subject of further work.

\subsection*{5.3 Is non-locality necessary}

In the previous two sections we have shown that  nonlocality still allows us to reproduce our "local" everyday experience. Even if this is true, one still can't help but feel that the local theory is aesthetically better. In fact, the technique of introducing Coulumb's law is generic, and it could have been used for non-relativistic quantum mechanics as well. Nevertheless, in non-relativistic case Bohm had instead chosen a "local" alternative. So this leads to a question: perhaps there is a way to come up with quasi-local flows that represent creation and annihilation of particles? If the answer is yes, we will have to justify why we haven't done it in this paper. We will make our argument by attempting to do what we are "not doing" and then shooting ourselves in a foot as we try. 

One thing that comes to mind is a paper by D\"urr et el (\cite{jumps}) where they have shown that a Pilot Wave model can be represented as a continuous evolution of fixed number of particles, interrupted by random "jumps" (which represent creation and annihilation of particles). The probability that a jump between state $\vert e >$  and $\vert e' >$ occurs within time interval $dt$ is $\sigma (e, e') dt$, where 
\beq \sigma (e, e') = \frac{<\psi \vert e> <e \vert H \vert e'> <e' \vert \psi>}{\vert <\psi \vert e > \vert^2}. \eeq
This means that in this paper we \emph{could} have used the concept of visibility (or something similar) to introduce a "continuous" version of that jump, and then turn that stochastic equation into deterministic one. While this is certainly a good thing to explore for a next project, there is a reason why we haven't done it here. Namely, in our work the particles are distinguishable while in the work of D\"urr et el they are not. Thus, we can't simply talk about jumps between two "states". We have to ask ourselves why would any individual particle "cooperate" with that jump. In light of the fact that particles have $0$ size, the probability of them running into each other is $0$. This means that the event of creation and annihilation has to occur over a short distance. Thus, the above "cooperation" is non-local. 

So far it is not so bad. The idea that particles have size is quite intuitive, and the above quasi-locality is consistent with that. But there is another problem: if we have more than just two particles, it is ambiguous which particle belongs to which pair. For example, if we have two electrons and one positron, \emph{which} of the electrons should annihilate a positron (once again, this question does not arise in the theory of D\"urr et el since in that work the particles are assumed to be indistinguishable)? Thus, we will have electron-electron interaction: if electron number $1$ has become invisible, it forces electron number $2$ to stay visible, and visa versa. Obviously, electron-electron interaction is not part of the paper by D\"urr et el, nor is there any vertex in quantum field theory to tell us what it is. So, we already have to do something more than trivial. 

Still, however, we are not in such a big trouble yet, since  the above electron-electron interaction is a quasi-local one; and, as we said before, quasi-local interactions are consistent with our intuition that particles have size. However, things become even more interesting if there is a chain of these interactions. For example, suppose there are $n$ electrons and $n$ positrons, positioned in such a way that electron number $k$ is supposed to annihilate positron number $k$. Now suppose we put a new electron (we will call it electron number $0$) right next to positron number $1$. Then, since positron number $1$ was annihilated by electron number $0$, the electron number $1$ has no choice but annihilate positron number $2$. Thus, electron $2$ will annihilate positron $3$, etc. Finally, electron number $n-1$ will annihilate positron $n$, while electron number $n$ will \emph{not} be annihilated -- we ran out of positrons! In other words, the presence of electron number $0$ \emph{caused} electron number $n$ to continue to stay visible. But, if $n$ is large enough, electron number $n$ might be several kilometers away from the electron number $0$. Thus, we have a real non-locality. 

There is still a way of dealing with it so far. In particular, we can simply allow the non-conservation of number of electron beables, which would remove a need of the above non-local interaction. After all, we never observed the electrons anyway (we only saw dots on the screen that were hit by electrons), so we have no direct evidence that their number is conserved. True, one evidence of their conservation is the fact that standard quantum field theory, which is based on that assumption, predicts correct results. This, however, is only an evidence of conservation in Hilbert space $Q$, but \emph{not} in beable space $B$. Suppose we have two states: one has extra electron and the other doesn't. \emph{Each} of these two states has been produced by  \emph{its own} quantum mechanical process that, of course, preserves the number of particles. While these two quantum processes are different, they are similar enough to stay within the same branch of a wave function in a configuration space. So we can freely move from one state, which is the outcome of the first process, to the second state, which is the outcome of the other one, without any detectable inconsistencies.

In order to truly get ourselves into trouble, we have to come up with a scenario where we have \emph{a lot of} electrons at stake; not just one. Suppose we have a very large number of electrons and positrons saturated within a lab, with a very large density. Furthermore, suppose we have measured their energy. Thus, by uncertainty principle, we have no information about their position. Quantum field theory predicts that if we will measure their position, it will be consistent with Poisson distribution. This will be true both before and after the annihilation events (up to the fact that the density will change). On the other hand, in our case we will \emph{not} expect Poisson distribution after the annihilation. After all, the particles that are closer to each other are more likely to annihilate. So, the ones that will be left will be further apart. Of course, in Poisson case we will also expect them to be further apart due to the smaller density. But in our case that effect will be even larger, since it will be combination of smaller density \emph{and} the "natural selection". 

At first one might think that this is not a big deal since the typical size between particles is so small that we will not detect their distance. But if our lab is large enough then the Poisson distribution will predict density fluctuations on large scales. That difference will be detectable. Furthermore, if the average distance between particles will be larger than the one predicted by lowering the density of Poisson process, this means that their rate of annihilation will be slowing down \emph{more} than expected from quantum field theory. Both of these phenomena can produce considerable inconsistencies with observations.  

Now, in order for the prediction of quantum field theory to be reproduced, we have to "make up" for it: whenever the annihilation process occurs, we have to  "move" the particles that are left in such a way that mimics Poisson distribution. If, however, the velocities of the particles are completely determined by "local" guidance equation, we will not be able to do that. After all, we have to "move" them based on non-local considerations. This is where "Coulumb's law" becomes very helpful! The latter, naturally, preserves divergence of the so-called "electric field". Thus, if the annihilation process "prefers" outcome $\vec{p} \in B$ over outcome $\vec{q} \in B$, then, in the "visibility" direction, the flow into $\vec{p}$ is larger than the one into $\vec{q}$. So, in order to the \emph{total} flows to be the same, there has to be a flow from $\vec{p}$ to $\vec{q}$ in some other direction. The only direction that is left is regular space direction. Thus, the latter flow does the moving around of particles for us. This can not be accomplished so easily without resorting to Coulumb's law.

One can still argue that high saturation of particles might be excluded through ultraviolet cutoff. But, in order for the annihilation of particles to occur often enough, the scale of the annihilation interaction has to have a lower bound. If we insist on using ultraviolet cutoff argument, this will introduce an upper bound on the latter. That might be the ultimate reason why we decided to use Coulumb's law in this paper. However, since our attempts to do other things were not completely unproductive, it might still be useful to explore these other approaches in future work. But, at the same time, we still have a good excuse for sticking with Coulumb's law at least temporarily. 

\subsection*{6. Comparison of my work to the one of Nikoli\'c}

As I have mentioned in the last paragraph of Introduction, Nikoli\'c  has proposed similar idea to mine (see \cite{nikolicb} and \cite{nikolicf}) reviewing his work and discussing similarities and differences between the concepts that we have proposed. 

The starting point of his work is the idea of \emph{field beables}, which was also used by others, including W. Struyve and H. Westman (\cite{minimalist}). Consider, for example, spin $0$ field. We can view second quantization of the latter as a \emph{first} quantization of oscillating lattice. Since the number of lattice points is conserved, we no longer have to worry about particle creation and annihilation. This should be obvious from the intution we have by observing water waves: while the latter can be created and destroyed, water molecules are \emph{not}. Thus, since a dynamics of water molecules is continuous and deterministic, the same applies to water waves, by default.

What distinguishes the work of Nikoli\'c from others, however, is that he drew a link \emph{back} from field to particles; by doing this he has shown that the concept of field beables dictates the concept of "effectivity" of the particles. Consider again the example of water waves. This time, however, let us \emph{insist} on writing the dynamics in terms of the waves and not water molecules (after all, Fourier analysis demands that this should be possible). Since we know that the water molecules are subject to differentiable laws, we also know that the Fourier transform of their dynamics is differentiable as well. In other words, the amplitudes of the waves can not go from their generic value to $0$ instantaneously; rather, there has to be a small transition period. That is precisely what I called "visibility" and what Nikoli\'c called "effectivity". In my case, visibility was simply postulated. In his case, effectivity was a \emph{prediction} made based on above argument. 

Let us now go back to quantum mechanical view to show more precisely what is going on. If we have a single field $\phi$, then a quantum state is a probablity amplitude distribution $\psi (\phi)$. If $\tilde{\phi}$ is a Fourier transformation of $\phi$ into a momentum space, we can define $\tilde{\psi}$ through 
\beq \tilde{\psi} (\tilde{\phi}) = \psi (\phi). \eeq
 It is easy to show that the lattice can be represented as multi-dimensional harmonic oscillator, where each dimension corresponds to a given momentum. In case of lack of perturbation, it decomposes by simple equation
\beq \tilde{\psi} (\tilde{\phi}) = \prod_{\vec{k}} A_{\omega_{\vec{k}} } (\tilde{\phi} (\vec{k})) e^{i \omega_{\vec{k}} (t - t_{\vec{k}})} \eeq
where $A_{\omega}$ represents the \emph{position} variable of \emph{one} of the harmonic oscillators with natural frequency $\omega$, $t_{\vec{k}}$ is an "innitial time" for any given oscillator, and 
\beq \omega_{\vec{k}} = \sqrt{ \vert \vec{k} \vert^2 + m^2} \eeq
 Now, any particular behavior of $\phi$ is in one to one correspondence with 
\beq \psi (\phi') = \delta (\phi' - \phi) \eeq
Furthermore, from the completeness of the set of solutions of Harmonic oscillator, we know that the latter can be \emph{uniquely} defined as a linear combination of \emph{all} of the solutions:
\beq \delta (\tilde{\psi}' (\vec{k}) - \tilde{\psi} (\vec{k}) ) = \sum \psi_{n, w_{\vec{k}}} (\tilde{\psi}' (\vec{k})) (a_{\vec{k}})^n \vert 0 > \eeq
Now, the complete description of $\tilde{\psi}$ is given by 
\beq \prod_{\vec{k}} \delta (\tilde{\psi}' (\vec{k}) - \tilde{\psi} (\vec{k}) ) = \prod_{\vec{k}} \Big( \sum \psi_{n, w_{\vec{k}}} (\tilde{\psi}' (\vec{k})) (a_{\vec{k}})^n \vert 0 > e^{i (t - t_{\vec{k}})}\Big) \eeq
The above correspondence is one to one but \emph{not} onto. In other words, to each $\phi$ corresponds a unique quantum state (which is only unique to that specific $\phi$), \emph{but} some quantum states do not correspond to any $\phi$ at all. In other words, by imposing field beables we are defining "mixed state" and \emph{also} imposing a relationship between any two coefficients of that state.

However, we are already used to reducing the number of degrees of freedom in case of particle beables. In the latter case, we again impose a relation between coefficients of different states: namely, one of the coefficients has to be $1$ and the rest should be $0$. By replacing particle beables with field ones, we are simply substitutting the latter for some other, more complicated, relatinoship. The presence of mixed states, however, leads to a concept of \emph{effectivity} of particles. In particular, the effectivity of \emph{pure} state $\vert s_k >$ is given by
\beq e(\vert s_k >) = \frac{ \vert < \psi \vert s_k> \vert^2}{\sum_l \vert < \psi \vert s_l> \vert^2} \eeq
which, clearly, is a number between $0$ and $1$, similar to visibility. When particles are viewed as beables, effectivity is forced to be either $0$ or $1$. When field beables are introduced, that is not the case; but still there is no freedom left in defining effectivity; it is defined for us through the well known solutions of Harmonic oscillator. This is one of the key characteristics that distinguishes Nikoli\'c proposal from mine. A lot of other differnces might be viewed as a logical consequences of this one.  Let us therefore list all the differences between our proposals more formally: 

1. With a risk of being redundant, we will repeat for formality's sake the difference I have just mentioned: According to Nikoli\'c the field should be treated as a beable, and then the behavior of the field is 'translated" into the effectivity of particle states. This leaves no freedom in terms of making sure that effectivity is close to $0$ and $1$ most of the time (or imposing any other conditions, for that matter). According to Nikoli\'c the latter should be a natural consequence of decoherence.  On the other hand, in terms of my proposal, the above criteria on visibility is \emph{forced} by means of differentiable approximation to step function (which I call $f$). While "forcing" something puts us on a "safer side" it also makes a theory a lot less natural. Thus, the ultimate answer to the question of whether mine theory is "better" or his lies in a validity of a conjecture that decoherence leads to a "split" along the lines characterized by particle numbers. While Nikoli\'c quotes some work that proves something to that effect, I haven't had time, yet, to study it. 

2. In light of the fact that in Nikoli\'c case the effectivity arizes naturally from standard quantum field theory, he does \emph{not} introduce any extra compactified coordinates to accomplish this goal. 

3. According to Nikoli\'c, the particles are indistinguisheable, while according to my proposal they are, in fact, dustinguisheable. The reason for the indistinguisheability of the particles in Nikoli\'c's case is that the latter can be argued for on a basis of harmonic oscillator: the number of particles is simply the excitation level of the latter. Thus, when harmonic oscillator is excited to a level $2=1+1$, it doesn't make sense to ask what is the difference between one $1$ and the other $1$. In my case, however, both particles, as well as their creation and annihilation was put by hand. That is what allowed them to be distinguisheable.

4. As a natural consequence of the fact that particles are indistinguisheable in Nikoli\'c's case, he was forced to impose the notion of the effectivity on the \emph{entire} state, rather than any particular particle. This made the latter non-local. In my case, however, non-locality is still very much present in the \emph{dynamics} of particles (as discussed in detail in Chapter 5). Thus, the only difference is that the \emph{definition} of visibility, \emph{apart from all dynamics}, is local. The importance of the latter is questionable. 

5. In his work he did not have to resort to bluntly non-local tricks, like Coulumb's law. I believe what helped him to avoid this is the fact that his theory reduces to field beables, and the latter can be equated with \emph{first} quantization of the lattice, with fixed number of points. Finally, we know from Bohm that the pilot wave models of \emph{first} quantization can be done without sacrifice of the locality in configuration space. But, Nikoli\'c has shown that the \emph{first} quantization of the lattice is in one to one correspondence with a \emph{second} quantization of a theory with varying "effectivity" degrees. Thus, logic implies that his "effectivity" theory also meets the desired property. This is something I was not able to do since in my case particles were postulated by hand, and thus were \emph{not} viewed in terms of quantum states, which ruined the above analogy.

6. In light of the fact that Fermionic field is Grassmann- valued, he was not able to use directly field beables for fermions, Instead, he "bosonified" the fermionic field by replacing anti-commutting creation and annihilation operators with the commutting ones. In my case, however, i was able to treat fermions and bosons on the same grounds, since I did not use field beables to begin with. However, the importance of this is questionable in light of the fact that, on a grand scheme of things, I had to use a lot more unnatural tricks. 

7. As with any other two different Pilot Wave models, it might be of interest to see whether changing the choice of beables would lead to different predictions. Roughtly speaking, in my case the particle position is being "continuously measured" while in his case the field is. The fact that one is a "transformation" of the other does not imply that the predictions of the theories should be identical. After all, momentum is a "transformation" of position; yet we know that measuring one of the two would produce very different phenomenology from measuring the other one. For the same reason, both my theory and Nikoli\'c's should be further investigated and any differences in predicted phenomenologies should be explored. 

\subsection*{7. Conclusions}

In this paper we have found how Bohm's Pilot Wave model can be supplemented with creation and annihilation of beables. By claiming that particles never get annihilated but instead ``become invisible", we were able to introduce the notion of non-instantaneous creation or annihilation by allowing the ``visibility" to vary between 0 and 1. This continuity allowed the creation and annihilation to be incorporated into the Pilot Wave model.  

In light of the fact that particles are neither created nor destroyed but simply change their visibilities, the total number of ``visible" particles is bounded above. Thus, our definition of amplitudes of extreme states (before the weighted averages were taken) is dependent on the effective field theory that predicts that probability having more than $N$ particles is very small. This can be accomplished by imposing a UV cutoff on a theory and taking that cutoff seriously as a real physical constant.  

One thing that is worth exploring is whether or not the continuously changing "visibility" of the particles can be replaced with something more physical, for example, a continuous "rotation" of electron into neutrino. The reason it was not done in this paper is that there are some processes that can not be depicted as rotations, such as for example a decay of photon into electron and positron pair.  

It should also be pointed out that this approach is not the only one that allows us to incorporate quantum field theory. For example, Struyve and Westman have developed a minimalist theory \cite{minimalist} where only bosonic beables were postulated. After all, one can argue that we didn't really see an electron; we only saw its electromagnetic field that told us where that ``electron" was supposedly located. Since bosons can be described through field beables (which is much harder to do for fermions due to their Grassmannian nature), this completely avoids the need to create or destroy particles.  

Even more importantly, Nikoli\'c' took the theory of field beables one step further and actually drew a connection from that concept to the concept of \emph{effectivity} of particle states (see \cite{nikolicb} and \cite{nikolicf}). The latter is very similar with the "visibility" proposed in my work (one key difference is that the former applies to multiparticle states, where particles are indistinguishable, while the latter applies to a \emph{single} particle, which \emph{is} distinguisheable). The similarities and differences between my proposal and his were discussed in Chapter 6. 

Finally, there is yet another alternative, proposed by Colin \cite{collin}, in which he retained the presence of fermions, but got rid of the concept of creation and annihilation in favor of taking the concept of Dirac sea seriously. Of course, this raises some conceptual questions, such as just how deep the Dirac sea is, and what are its implications in terms of vacuum energy.  I, personally, believe that all such models should be explored in parallel.

\end{document}